\documentclass[twocolumn]{aastex63}
\def\actaa{Acta Astronomica}
\def\feh{$\mathrm{[Fe/H]}$}

\usepackage{amsmath}

\hypersetup{linkcolor=red,citecolor=blue,filecolor=green,urlcolor=magenta}

\begin{document}

\shorttitle{Segue II \& Ursa Major II RR Lyrae}
\shortauthors{Ngeow \& Bhardwaj}

\title{Photometric Metallicity and Distance for the Two RR Lyrae in Segue II and Ursa Major II Dwarf Galaxies Based on Multi-Band Light-Curves}

\correspondingauthor{Chow-Choong Ngeow}
\email{cngeow@astro.ncu.edu.tw}

\author[0000-0001-8771-7554]{Chow-Choong Ngeow}
\affil{Graduate Institute of Astronomy, National Central University, 300 Jhongda Road, 32001 Jhongli, Taiwan}

\author[0000-0001-6147-3360]{Anupam Bhardwaj}
\affil{INAF-Osservatorio astronomico di Capodimonte, Via Moiariello 16, 80131 Napoli, Italy}

\begin{abstract}

Multi-band light curves of two RR Lyrae variables in Segue II and Ursa Major II ultra-faint dwarf (UDF) galaxies were collected from near simultaneous observations using the Lulin One-meter Telescope in $Vgri$ bands. Together with Gaia $G$-band light curves, we determined photometric metallicities using empirical relations involving pulsation period and Fourier parameter as dependent parameters. We demonstrated that the RR Lyrae photometric metallicity can be determined accurately when these empirical relations were employed at multiple wavelengths, which can potentially improve the distance determination based on RR Lyrae stars. The photometric metallicities based on our approach were found to be $-2.27 \pm 0.13$~dex and $-1.87 \pm 0.16$~dex for the RR Lyrae in Segue II and Ursa Major II UFD, respectively, with corresponding distance moduli of $17.69 \pm 0.15$~mag and $17.58 \pm 0.15$~mag, in agreement with previous literature determinations. This approach of photometric metallicity of RR Lyrae based on multi-band optical light curves will be particularly relevant for distance measurements in the era of the Vera C Rubin’s Legacy Survey of Space and Time. 

\end{abstract}


\section{Introduction}\label{sec1}

The old-population pulsating stars RR Lyrae are well-known distance indicators because they exhibit an absolute $V$-band magnitude-metallicity relation and well-defined period-luminosity(-metallicity) relations in other filters \citep[for a review, see][]{bhardwaj2020}. In most cases, the metallicity of RR Lyrae, parameterized as \feh, needs to be known a priory to apply these relations. The best \feh~measurements are based on (high-resolution) spectroscopic observations. However, such observations could be time-consuming and are only limited to relatively bright RR Lyrae in the Milky Way. Alternatively, \feh~can be obtained using the light-curve information, as shown in the seminar paper by \citet{jk1996}. The \feh~measured using this approach, known as photometric metallicity, is expected to be widely applied to the distant RR Lyrae discovered from the multi-band time-domain sky surveys, especially the Vera C. Rubin Observatory's Legacy Survey of Space and Time \citep[LSST,][]{lsst2019}, because only few spectrographs on large-aperture telescopes, or perhaps none, can be used to collect high-resolution spectra of very distant RR Lyrae stars.  

In the past, RR Lyrae photometric metallicities were typically obtained using light curve structure information in a single filter. On the other hand, empirical relations that use the light-curve information to estimate \feh~have been derived in several filters, ranging from optical $V$-band to infrared WISE filters (some examples can be found in Section \ref{sec2}). Higher precision photometric metallicity can be achieved by averaging out the \feh~obtained from several such relations in the same or different filters mitigating possible systematic effects. In this work, we demonstrate this is indeed the case based on the multi-band observations of two ab-type (i.e. fundamental-mode) RR Lyrae discovered in the ultra-faint dwarf galaxies SEGUE II \citep{boettcher2013} and Ursa Major II \citep[hereafter UMaII,][]{dallora2012}. These near-simultaneous time-series observations were carried out using the Lulin One-meter Telescope (LOT),\footnote{We note that \citet{vivas2020} discovered three additional RR Lyrae associated with UMaII. However, these three RR Lyrae were located outside the field-of-view (FOV) for LOT. In contrast, no additional RR Lyrae were found for SEGUE II \citep{vivas2020}.} a general-purpose Cassegrain reflector located at the Lulin Observatory in central Taiwan. The collected data allow us to obtain the photometric metallicities for these two RR Lyrae based on the homogeneous light curve data at multiple wavelengths.

Section \ref{sec2} presents multi-band light curves collected from LOT, together with archival Gaia light curves, for these two RR Lyrae. Based on these multi-band light curves, we obtained their photometric metallicity in Section \ref{sec3}. A by-product of our work is the derivation of distance moduli to these dwarf galaxies, which will be discussed in Section \ref{sec4}. The conclusions of our work is presented in Section \ref{sec5}. Throughout this work, these two RR Lyrae will be referred as SEGUE II-V1 and UMaII-V1, respectively.

\section{Multi-Band Light Curves} \label{sec2}

\begin{figure*}
  \epsscale{1.2}
  \plotone{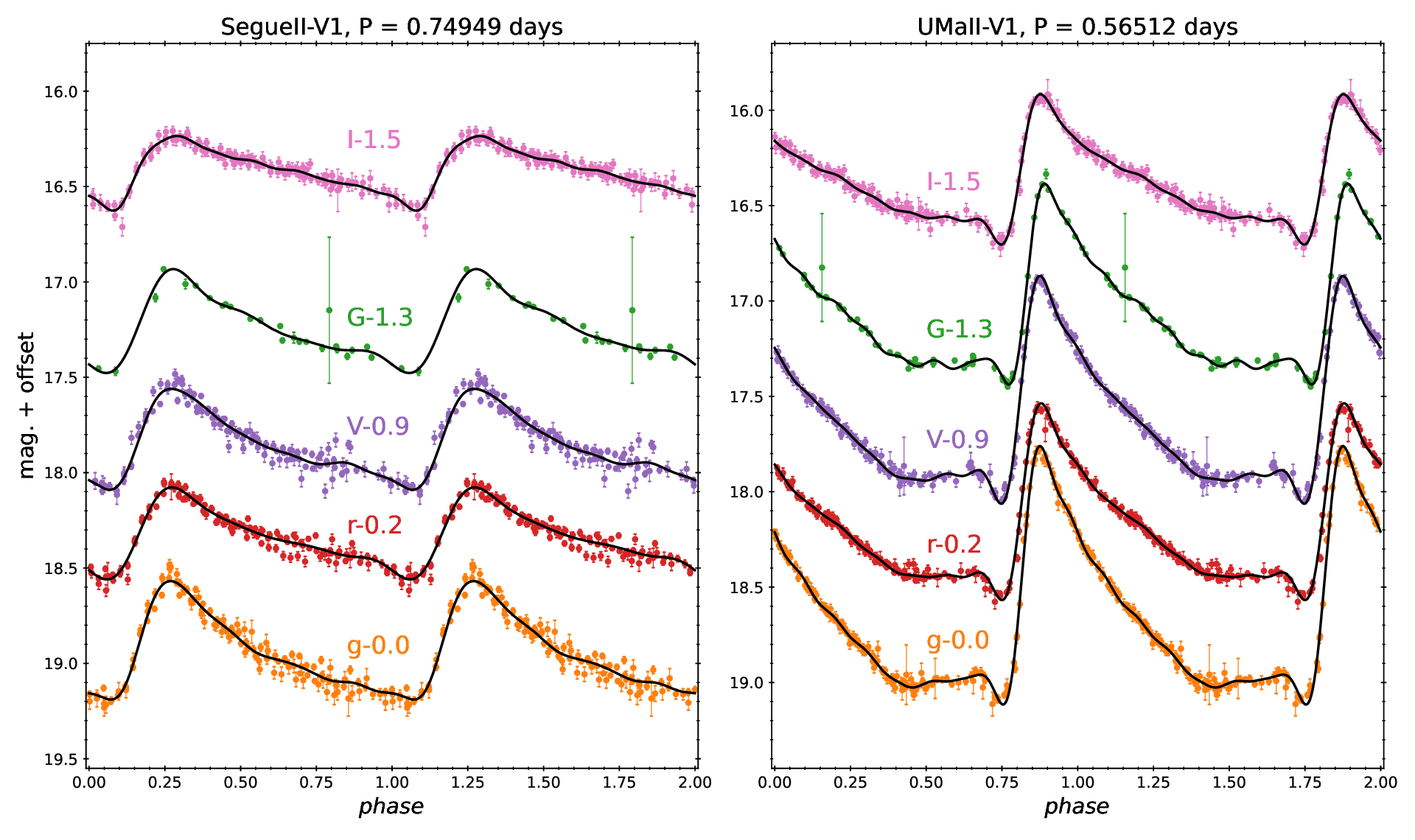}
  \caption{The $grVI$-band light curves based on LOT observations and the $G$-band light curves retrieved from Gaia DR3 for SEGUE II-V1 (left panel) and UMaII-V1 (right panel). For a better visualization, these light curves, except the $g$-band, were offsetted. The black curves represent the best-fit light curves based on the Fourier-expansion.}
  \label{fig_2lc}
\end{figure*}

Time-series observations of SEGUE II-V1 and UMaII-V1 were carried out with LOT in 12 and 18 nights, respectively, between 2019 November 14 and 2020 April 15. The Princeton Instruments SOPHIA 2048B back-illuminated CCD was mounted on LOT during these observations, providing a FOV of $13.2\arcmin \times 13.2\arcmin$ (with a pixel scale of $0.386\arcsec$/pixel) on the CCD images. A sequence of $Vgri$ exposures, with identical exposure time of 360~second in all filters, were taken multiple times for each RR Lyrae in a given night when they were visible from the Lulin Observatory. All of the collected images were reduced in a standard manner (including bias- and dark-subtraction, followed by flat-fielding). Astrometric calibration on the reduced images were done using the {\tt SCAMP} package \citep{scamp2006}. The PSF (point-spread function) photometry of RR Lyrae in the reduced images were measured using the combined {\tt PSFEx} \citep{bertin2011} and {\tt Source-Extractor} \citep{bertin1996} package, at which the details on the photometric calibration were presented in the Appendix A. Table \ref{tab1} and \ref{tab2} provide multi-band light curves collected from the LOT observations for SEGUE II-V1 and UMaII-V1, respectively.

\begin{deluxetable}{cccc}
  \tabletypesize{\scriptsize}
  \tablecaption{Light Curves for SEGUE II-V1.\label{tab1}}
  \tablewidth{0pt}
  \tablehead{
    \colhead{$JD$} &
    \colhead{mag} &
    \colhead{mag\_err} &
    \colhead{Filter} 
  }
  \startdata
2458831.0928240740 & 19.150 & 0.042 & g \\
2458839.9920949074 & 18.961 & 0.014 & g \\
2458802.1246759258 & 18.824 & 0.025 & g \\
2458831.1647916664 & 19.078 & 0.051 & g \\
2458830.0386458333 & 18.798 & 0.026 & g \\
$\cdots$ & $\cdots$ & $\cdots$ & $\cdots$ \\
\enddata
\tablecomments{This table is available in its entirety in machine-readable form, a portion is shown here for guidance on its content.}
\end{deluxetable}

\begin{deluxetable}{cccc}
  \tabletypesize{\scriptsize}
  \tablecaption{Light Curves for UMaII-V1.\label{tab2}}
  \tablewidth{0pt}
  \tablehead{
    \colhead{$JD$} &
    \colhead{mag} &
    \colhead{mag\_err} &
    \colhead{Filter} 
  }
  \startdata
2458911.1549074072 & 18.298 & 0.017 & g \\
2458909.9914583331 & 18.119 & 0.029 & g \\
2458931.0296990741 & 18.653 & 0.024 & g \\
2458850.3490856481 & 18.959 & 0.156 & g \\
2458885.1327662035 & 18.153 & 0.019 & g \\
$\cdots$ & $\cdots$ & $\cdots$ & $\cdots$ \\
\enddata
\tablecomments{This table is available in its entirety in machine-readable form, a portion is shown here for guidance on its content.}
\end{deluxetable}

In addition to the new LOT observations, we have also downloaded $G$-band light curves released from the Gaia Data Release 3 \citep[DR3,][]{gaia2016,gaia2023} for these two RR Lyrae, with Gaia DR3 source ID for SEGUE II-V1 and UMaII-V1 as 87207528534363264 and 1043841876592990208, respectively. We have also adopted their periods as derived in Gaia DR3 \citep[][from the {\tt I/358/vrrlyr} Table available in the SIMBAD/VizieR database]{clementini2023,eyer2023}, because the adopted periods are adequate to fold the light curves and there is no need to re-derive their periods. The adopted periods are $P(\mathrm{SEGUE\ II}$-$\mathrm{V1}) = 0.7494938\pm0.0000164$~days and $P(\mathrm{UMaII}$-$\mathrm{V1}) = 0.5651226\pm0.0000037$~days.

Figure \ref{fig_2lc} presents multi-band phase folded light curves with the adopted periods. These light curves were then fitted with a low-order Fourier-expansion \citep[see][and reference therein]{ngeow2022}. In brief, we fitted a Fourier series of the following form:

\begin{eqnarray}
  m(\Phi) & = & m_0 + \sum^n_{i=1} A_i \sin (2 \pi i \Phi + \phi_i), \nonumber
\end{eqnarray}

\noindent to the phased light curves, where $\Phi$ is pulsational phase from 0 to 1, and Fourier order $n$ varied from 4 to 9. The best order of fit was determined based on the method as described in \citet{ngeow2023}. From the best-fitted Fourier expansions, we determined the $\phi_{31}=\phi_3-3\phi_1$ Fourier parameters which are needed to estimate the photometric metallicity for RR Lyrae (see next Section). The determined $\phi_{31}$ from the multi-band light curves are provided in Table \ref{tab3}. 

\section{The Metallicity of RR Lyrae} \label{sec3}

\subsection{The Adopted Relations}

It is well-known that \feh~of a RR Lyrae can be obtained from its pulsation period $P$ and its light curve parameter, primarily the Fourier parameter $\phi_{31}$ \citep{jk1996}. We collected the $grGVI$-band $[\mathrm{Fe/H}]$-$\phi_{31}$-$P$ relations that are available recently in the literature. These include the $grV$-band relations taken from \citet[][hereafter N22]{ngeow2022}:

\begin{eqnarray}
  [\mathrm{Fe/H}]^g_{N22} & = & -1.01[\pm0.06] - 7.43[\pm0.80](P-0.58) \nonumber \\
  & & +1.69[\pm0.16](\phi_{31}-5.25), \ \ \sigma = 0.24,  \nonumber
\end{eqnarray}
\begin{eqnarray}
  [\mathrm{Fe/H}]^r_{N22} & = & -1.54[\pm0.07] - 8.28[\pm1.04](P-0.58) \nonumber \\
  & & +1.30[\pm0.15](\phi_{31}-5.25), \ \ \sigma = 0.30,  \nonumber
\end{eqnarray}
\begin{eqnarray}  
  [\mathrm{Fe/H}]^V_{N22} & = & -1.21[\pm0.05] - 7.67[\pm0.78](P-0.58) \nonumber \\
  &  & +1.50[\pm0.14](\phi_{31}-5.25), \ \ \sigma = 0.24.  \nonumber
\end{eqnarray}

\noindent These relations were derived based on a sample of 30 RR Lyrae located in the Kepler Field, and their spectroscopic $[\mathrm{Fe/H}]$ were derived from high-resolution spectra \citep{nemec2013}. The metallicity scale for N22, $[\mathrm{Fe/H}]_{N22}$, is in the \citet{carretta2009} scale. We have also adopted $V$-band $[\mathrm{Fe/H}]$-$\phi_{31}$-$P$ relation from \citet[][hereafter M21]{mullen2021}, derived from 1980 field RR Lyrae, as:

\begin{eqnarray}
  [\mathrm{Fe/H}]^V_{M21} & = & -1.22[\pm0.01] - 7.60[\pm0.24](P-0.58) \nonumber \\
  &  & +1.42[\pm0.05](\phi_{31}-5.25), \ \ \sigma = 0.41.  \nonumber
\end{eqnarray}

\noindent The M21 metallicity is in \citet{crestani2021} scale, therefore we applied an offset of $-0.08$~dex to convert $[\mathrm{Fe/H}]^V_{M21}$ to the \citet{carretta2009} scale as determined in \citet{mullen2021}. We did not include $V$-band relation from \citet{jk1996} because the qualities for some of the light curves used in their work \citep[as commented in][]{mullen2021,zong2023} are not as good as those studied either in M21 or N22.

In the Gaia $G$-band, \citet[][hereafter I21]{iorio2021} derived the relation using a set of 86 field RR Lyrae, as given below:

\begin{eqnarray}
  [\mathrm{Fe/H}]^G_{I21} & = & -1.68[\pm0.05] - 5.08[\pm0.5](P-0.60) \nonumber \\
   & & +0.68[\pm0.11](\phi_{31}-2.00-\pi),  \ \ \sigma = 0.31.  \nonumber
\end{eqnarray}

\noindent \citet{mullen2021} pointed out that a $\pi$ is needed to be subtracted from $\phi_{31}$ when using their relation, hence we included it in the above relation. The metallicity scale for the I21 relation is in \citet{zinn1984} scale, therefore we converted the \citet[][ZW84]{zinn1984} scale to the \citet[][C09]{carretta2009} scale using the relation given in \citet{carretta2009}: $[\mathrm{Fe/H}]_{C09} = 1.105[\mathrm{Fe/H}]_{ZW84}+0.160$. Independently, \citet[][hereafter J23]{jurcsik2023} derived the $G$-band relation using RR Lyrae-hosted globular clusters. Since it is well-known that globular clusters can be divided into Oosterhoff-I (OoI, with higher metallicity) and Oosterhoff-II (OoII, with lower metallicity) types, \citet{jurcsik2023} derived three sets of $G$-band relation: one for each Oosterhoff types and a relation for the combined sample. Since both SEGUE II-V1 and UMaII-V1 are OoII RR Lyrae \citep{boettcher2013,dallora2012}, we adopted the OoII relation:

\begin{eqnarray}
  [\mathrm{Fe/H}]^G_{J23} & = & -4.324[\pm0.157] - 7.212[\pm0.248]P \nonumber \\
   & & +1.353[\pm0.048]\phi_{31}, \ \ \sigma = 0.17.  \nonumber
\end{eqnarray}

\noindent As \citet{jurcsik2023} adopted a slightly different Solar compositions, and offset of $-0.04$~dex was added to $[\mathrm{Fe/H}]^G_{J23}$ to bring the metallicity to the \citet{carretta2009} scale. Besides these two $G$-band relations, another relation is available from \citet[][hereafter L23]{li2023}, which includes an additional $R_{21}=A_2/A_1$ Fourier parameter:

\begin{eqnarray}
  [\mathrm{Fe/H}]^G_{L23} & = & -1.888[\pm0.002] - 5.772[\pm0.026](P-0.6) \nonumber \\
  & & +1.090[\pm0.005](\phi_{31}-2.0-\pi) \nonumber \\
  & & +1.065[\pm0.030](R_{21}-0.45),  \ \ \sigma = 0.24.  \nonumber
\end{eqnarray}

\noindent The $R_{21}$ values from the $G$-band light curve are $0.495\pm0.022$ and $0.423\pm0.006$ for SEGUE II-V1 and UMaII-V1, respectively. Similar to the I21 relation, a $\pi$ is needed to be subtracted from the $\phi_{31}$ value. The above relation was derived from a sample of $\sim 2000$ RR Lyrae with spectroscopic $[\mathrm{Fe/H}]$ measured from \citet{liu2020}. According to \citet{liu2020}, the difference between their $[\mathrm{Fe/H}]$ and published $[\mathrm{Fe/H}]$, in the \citet{carretta2009} scale, is negligible. Hence, we assume $[\mathrm{Fe/H}]^G_{L23}$ is in the \citet{carretta2009} scale.

Finally, the $I$-band $[\mathrm{Fe/H}]$-$\phi_{31}$-$P$ relations were published in \citet[][hereafter S05]{smolec2005} and \citet[][hereafter D21]{dekany2021}. S05 provides a two-parameter and a three-parameter relation. However, \citet{hajdu2018} demonstrated that the two-parameter relation suffers from a systematic bias, therefore we adopted the three-parameter relation (which also has a smaller dispersion):

\begin{eqnarray}
  [\mathrm{Fe/H}]^I_{S05} & = & -6.125[\pm0.832] - 4.795[\pm0.285]P + 1.181\nonumber \\
  & & [\pm0.113]\phi_{31} + 7.876[\pm1.776]A_2, \ \sigma = 0.14. \nonumber
\end{eqnarray}

\noindent The metallicity scale for S05 relation is in the \citet[][J95]{jurcsik1995} scale, and we converted it to the \citet{carretta2009} scale using $[\mathrm{Fe/H}]_{C09} = 1.001[\mathrm{Fe/H}]_{J95}-0.112$ \citep{kapakos2011}. \citet{dekany2021} performed a feature selection and found that the optimal relation could also be obtained with three parameters, and subsequently applied a Bayesian linear regression on a sample of $\sim 80$ RR Lyrae with good quality $I$-band light curves to derive the following relation, which is similar to the S05 relation:

\begin{eqnarray}
  [\mathrm{Fe/H}]^I_{D21} & = & -5.819[\pm0.149] - 6.350[\pm0.067]P + 1.248\nonumber \\
  & & [\pm0.020]\phi_{31} + 5.785[\pm0.320]A_2, \ \sigma = 0.20. \nonumber
\end{eqnarray}

\noindent According to \citet{dekany2021}, the D21 metallicity scale is in, or has been shifted to, the \citet{crestani2021} scale. Hence, we applied the same offset of $-0.08$~dex as in the case of M21 relation to bring the D21 metallicity scale to the \citet{carretta2009} scale. For SEGUE II-V1 and UMaII-V1, the $A_2$ parameters in the $I$-band are $0.069\pm0.003$ and $0.114\pm0.003$, respectively.

\begin{deluxetable}{lcccc}
  \tabletypesize{\scriptsize}
  \tablecaption{The determined $\phi_{31}$ Fourier parameters and photometric metallicities for SEGUE II-V1 and UMaII-V1.\label{tab3}}
  \tablewidth{0pt}
  \tablehead{
    \colhead{Filter /} & 
    \multicolumn{2}{c}{SEGUE II-V1} &
    \multicolumn{2}{c}{UMaII-V1} \\
    \colhead{Relation} &
    \colhead{$\phi_{31}$} &
    \colhead{$[\mathrm{Fe/H}]$} & 
    \colhead{$\phi_{31}$} &
    \colhead{$[\mathrm{Fe/H}]$}  
  }
  \startdata
  $g$/N22 & $5.170\pm0.057$ & $-2.40\pm0.30$ & $4.703\pm0.024$ & $-1.82\pm0.27$ \\ 
  $V$/N22 & $5.491\pm0.060$ & $-2.15\pm0.29$ & $4.811\pm0.028$ & $-1.75\pm0.26$ \\ 
  $V$/M21 & $\cdots$        & $-2.25\pm0.42$ & $\cdots$        & $-1.81\pm0.41$ \\ 
  $G$/I21 & $5.615\pm0.091$ & $-2.18\pm0.33$ & $4.908\pm0.030$ & $-1.68\pm0.32$ \\ 
  $G$/J23 & $\cdots$        & $-2.17\pm0.42$ & $\cdots$        & $-1.80\pm0.36$ \\
  $G$/L23 & $\cdots$        & $-2.19\pm0.26$ & $\cdots$        & $-1.97\pm0.24$ \\
  $r$/N22 & $5.678\pm0.056$ & $-2.39\pm0.37$ & $4.949\pm0.027$ & $-1.81\pm0.31$ \\
  $I$/S05 & $6.212\pm0.114$ & $-1.95\pm1.13$ & $5.304\pm0.056$ & $-1.79\pm1.07$ \\
  $I$/D21 & $\cdots$        & $-2.51\pm0.32$ & $\cdots$        & $-2.21\pm0.29$ \\
\enddata
\end{deluxetable}

\subsection{The Derived Metallicity}

\begin{figure*}
  \epsscale{1.15}
  \plottwo{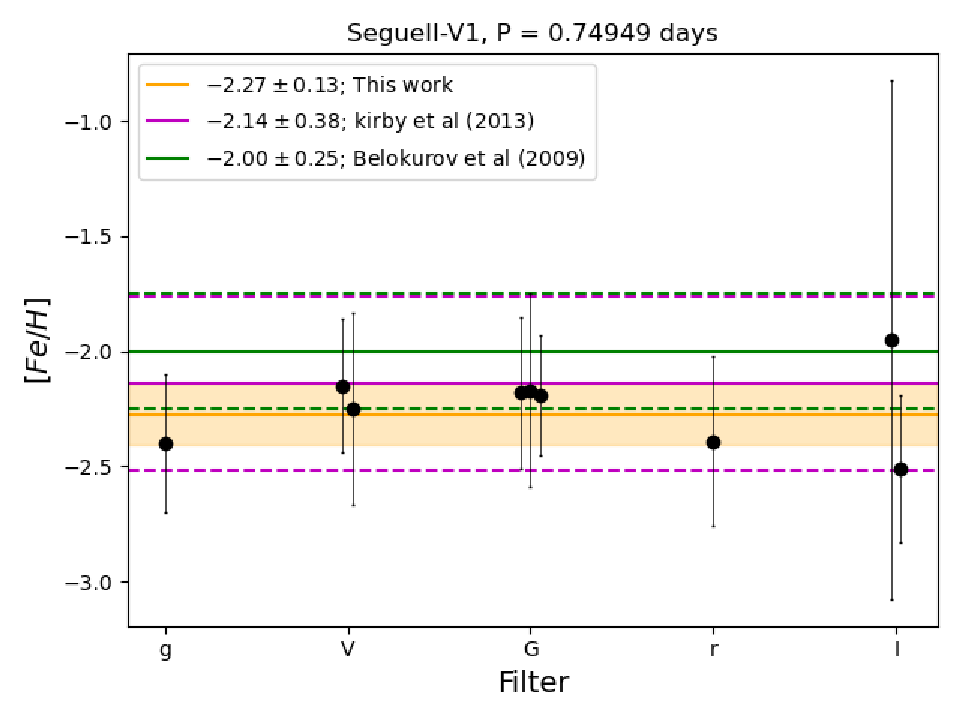}{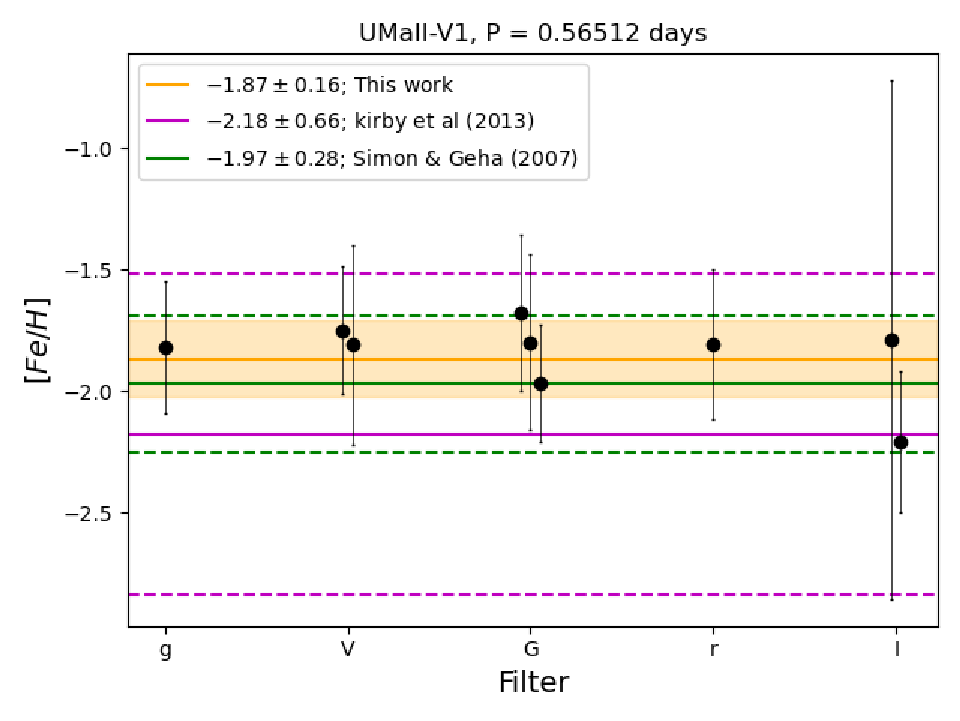}
  \caption{The derived photometric metallicities in various filters, as summarized in Table \ref{tab3}, for SEGUE II-V1 (left panel) and UMaII-V1 (right panel). The orange horizontal lines represent the weighted means based on the nine adopted relations, and the orange shaded regions are the $\pm1\sigma$ dispersion of the weighted means. Horizontal lines in green and magenta colors are the published values, together with their $\pm1\sigma$ dispersion shown as dashed lines, based on spectroscopic observations of a number of bright member stars in these two dwarf galaxies. For a better visualization, we only plotted the values taken from two publications for these two RR Lyrae.}
  \label{fig_feh}
\end{figure*}

The derived multi-band photometric metallicities, using the nine relations listed in the previous subsection, are listed in Table \ref{tab3} and graphically presented in Figure \ref{fig_feh}. As mentioned, all of these photometric metallicities are in the \citet{carretta2009} scale. For both RR Lyrae, all of the derived photometric metallicities are mutually consistent with each others from the nine relations adopted in the previous subsection. The errors on the derived metallicities are the quadrature sum of the propagated errors of each parameters (except we omitted the errors on period as they are negligible) and the dispersion $\sigma$ for each relations. The large errors on the $[\mathrm{Fe/H}]^I_{S05}$ values are mainly due to the large error on the constant term, $\pm0.832$, in the S05 relation. Errors from other relations are around $\sim 0.3$ to $\sim0.4$~dex.

We assume that the adopted nine relations in the previous subsection are equally well calibrated, although each relation may have their own systematic uncertainties \citep{jurcsik2023}. The uncertainties on the coefficients of these relations were propagated to the derived photometric metallicities. Therefore, we took the weighted means for the photometric metallicities listed in Table \ref{tab3} giving smaller weights to more uncertain measurements. When taking a weighted mean, we obtained $[\mathrm{Fe/H}]=-2.27\pm0.11$~dex and $-1.87\pm0.10$~dex for SEGUE II-V1 and UMaII-V1, respectively. Dispersions around these mean values were $0.13$~dex and $0.16$~dex, respectively, and will be adopted as the errors on these weighted means, which are $\sim 2$ to $\sim 3$ times smaller that those listed in Table \ref{tab3} (except for the S05 relation). We have also tried to eliminate certain data points, such as excluding those based on S05 relation or with the largest deviations, the resulted weight means do not differ by more than $\pm0.03$~dex from our preferred values. Small statistics of two RR Lyrae precludes us from quantifing the optimal number of relations or the preferred relation for accurate photometric metallicity determinations.

In case of SEGUE II-V1, our derived photometric metallicity is in good agreement with the values determined in \citet{belokurov2009} and \citet{kirby2013}, as shown in the left panel of Figure \ref{fig_feh}, and almost identical to the value recalculated by \citet[][$-2.26\pm0.14$~dex]{boettcher2013}. For UMaII-V1, our photometric metallicity is close to the value given by \citet{simon2007}, but more metal rich than those determined in \citet{kirby2013} or \citet[][$-2.36$~dex with $\sigma=0.56$]{willman2012}. However, dwarf galaxies are known to exhibit a wide spread in metallicity \citep[for examples, see][and reference therein]{simon2007,willman2012,kirby2013}. The published metallicities are mostly measured from the spectroscopic observations of red giant branch stars and/or blue horizontal branch stars, and hence they may not necessarily represent the photometric metallicities for RR Lyrae.

\section{The Distances of the Two RR Lyrae} \label{sec4}

\begin{deluxetable}{lcccc}
  \tabletypesize{\scriptsize}
  \tablecaption{Mean Magnitudes for SEGUE II-V1 and UMaII-V1.\label{tab4}}
  \tablewidth{0pt}
  \tablehead{
    \colhead{RR Lyrae} &
    \colhead{$\langle V \rangle$} &
    \colhead{$\langle I \rangle$} &
    \colhead{$\langle g \rangle$} &
    \colhead{$\langle r \rangle$} 
  }
  \startdata
  SEGUE II-V1 & 18.74 & 17.91 & 18.93 & 18.53 \\
  UMaII-V1    & 18.48 & 17.86 & 18.59 & 18.36 \\
  \enddata
  \tablecomments{An error of $0.02$~mag is adopted on these mean magnitudes.}
\end{deluxetable}

Since the Wesenheit magnitude, denoted as $W$, is reddening-free by construction \citep[see the Appendix in][]{madore1991,madore2023}, we adopted the recent period-Wesenheit-metallicity (PWZ) relations available in the literature to derive the distances to these two RR Lyrae with our photometric metallicities determined in the previous Section. For our $VIgr$-band light curves, the number of data points per light curves is more than 100, hence according to \citet{ngeow2022a} the errors on the mean magnitudes derived from these light curves are negligible. Nevertheless, we adopted a conservative error of $0.02$~mag on the intensity-averaged mean magnitudes derived from the fitted Fourier-expansion (see Figure \ref{fig_2lc}) on our multi-band light curves. The $VIgr$-band mean magnitudes are listed in Table \ref{tab4}.

\citet{neeley2019} calibrated multi-band PWZ relations based on the 55 Milky Way RR Lyrae together with their Gaia DR2 parallaxes. According to \citet{neeley2019}, metallicities for these 55 RR Lyrae are assumed to be in the \citet{zinn1984} scale. Therefore, we converted our derived metallicities to the \citet{zinn1984} scale by inverting the $[\mathrm{Fe/H}]_{C09}$-$[\mathrm{Fe/H}]_{ZW}$ conversion. After applying the $VI$-band PWZ relation from \citet{neeley2019}, we determined the $VI$-band based distance moduli ($\mu_{VI}$) to SEGUE II-V1 and UMaII-V1 as $\mu_{VI}\mathrm{(SEGUE\ II}$-$\mathrm{V1)}= 17.67\pm 0.20$~mag and $\mu_{VI}\mathrm{(UMaII}$-$\mathrm{V1)}= 17.56\pm 0.19$~mag.

As an independent check, we have also obtained the distance moduli for the two RR Lyrae using the $gr$-band PWZ relation as derived in \citet[][using the RR Lyrae in globular clusters]{ngeow2022a}, where the metallicity scale is in the \citet[][D16]{dias2016} scale. The conversion of such metallicity scale to the \citet{carretta2009} scale is given as $[\mathrm{Fe/H}]_{C09} = 0.99[\mathrm{Fe/H}]_{D16} - 0.05$ \citep{dias2016}. After we inverted our derived metallicity to the \citet{dias2016} scale, we obtained $gr$-band based distance moduli ($\mu_{gr}$) as $\mu_{gr}\mathrm{(SEGUE\ II}$-$\mathrm{V1)}= 17.71\pm 0.24$~mag and $\mu_{gr}\mathrm{(UMaII}$-$\mathrm{V1)}= 17.61\pm 0.24$~mag. The derived distance moduli based on the $VI$-band and the $gr$-band PWZ relations are summarized in Table \ref{tab5}.

\begin{deluxetable}{lcc}
  \tabletypesize{\scriptsize}
  \tablecaption{Derived Distance Moduli for SEGUE II-V1 and UMaII-V1.\label{tab5}}
  \tablewidth{0pt}
  \tablehead{
    \colhead{$\mu$ (mag.)} &
    \colhead{SEGUE II-V1} &
    \colhead{UMaII-V1}
  }
  \startdata
  $\mu_{VI}$   & $17.67\pm0.20$ & $17.56\pm 0.19$ \\
  $\mu_{gr}$   & $17.71\pm0.24$ & $17.61\pm 0.24$ \\
  \hline
  $\langle \mu \rangle$ & $\mathbf{17.69\pm 0.15}$ & $\mathbf{17.58\pm 0.15}$ \\ 
\enddata
\end{deluxetable}

It is worth to mention that the impact on the determined distance moduli is negligible ($\pm0.01$~mag at most) if we did not convert the metallicity to the \citet{zinn1984} or \citet{dias2016} scale. This is mainly due to the small metallicity terms (0.13 and 0.05~dex/mag, respectively) in the adopted PWZ relations. Since both of the $\mu_{VI}$ and $\mu_{gr}$ were determined from independent PWZ relations and in different filters, the averages of $\mu_{VI}$ and $\mu_{gr}$ were adopted as the final distance moduli for these two RR Lyrae. For SEGUE II-V1, we obtained $\langle \mu\rangle=17.69\pm 0.15$~mag, which is almost identical to the value found in \citet[][$17.7\pm 0.1$~mag]{belokurov2009} and \citet[][$D=34.4\pm2.6$~kpc or $\mu=17.68^{+0.16}_{-0.17}$~mag]{boettcher2013} measured from the blue horizontal-branch stars. Similarly, we obtained $\langle \mu\rangle=17.58\pm 0.15$~mag for UMaII-V1, and it is in good agreement with the estimated value given by \citet[][$\sim 17.5\pm 0.3$~mag]{zucker2006}, based on the isochrone fitting, and is almost the same as the one determined in \citet[][$17.60\pm 0.20$~mag]{vivas2020} when including additional RR Lyrae for UMaII. We also found statistically consistent distance measurements to these two dwarf galaxies if independent theoretical PWZ relations from \citet{marconi2015} and \citet{marconi2022} were adopted instead of the empirical relations. 

\section{Conclusions} \label{sec5}

In this work, we obtained homogeneous $grVI$-band light curves using LOT for two RR Lyrae in SEGUE II and UMaII ultra-faint dwarf galaxies. Together with their Gaia $G$-band light curves, we derived the photometric metallicities for these two RR Lyrae using nine $[\mathrm{Fe/H}]$-$\phi_{31}$-$P$ relations (some of them include an additional Fourier parameter) recently published in the literature. Hence, the weighted means of the photometric metallicities for SEGUE II-V1 and UMaII-V1 were found to be $[\mathrm{Fe/H}]=-2.27\pm0.13$~dex and $-1.87\pm0.16$~dex, respectively. Using these photometric metallicities, together with empirical PWZ relations, we determined the distance moduli to these two RR Lyrae as $\mu\mathrm{(SEGUE\ II}$-$\mathrm{V1)}= 17.69\pm0.15$~mag and $\mu\mathrm{(UMaII}$-$\mathrm{V1)}= 17.58\pm0.15$~mag. These distance and photometric metallicity measurements are in good agreement with previous determinations in the literature.

As mentioned in the Introduction, the motivation of this work is to demonstrate that a higher precision on photometric metallicity for RR Lyrae can be acheived when averaging the values derived from multiple metallicity-Fourier parameter(s)-period relations. This is indeed the case for the two RR Lyrae, SEGUE II-V1 and UMaII-V1, investigated in this work. Our proposed approach will be particularly important for various ongoing and upcoming multi-band time-domain sky surveys providing well-sampled RR Lyrae light curves in the Milky Way. This is especially true for LSST which will provide unprecedentedly large samples of distant RR Lyrae beyond our Galaxy and particularly in the LSST-discovered new dwarf galaxies. The spectroscopic follow-up observations of these variables will not be feasible either because they are too far/faint or due to lack of rather competitive observational time. In this scenario, photometric metallities for RR Lyrae can be determined using the proposed approach, since LSST will provide multi-filter light curves of RR Lyrae variables. However, it is critically important to adopt a homogeneous metallicity scale or homogenize metallicity measurements based on different empirical relation for better constrain on the photometric metallicities, as shown in detail in the manuscript. These photometric metallicities can then be used to derive the RR Lyrae-based distances, as well as to compare and/or constraint the inferred metallicity based on isochrone fitting to the color-magnitude diagrams of the newly-discovered dwarf galaxies. The large number of expected RR Lyrae in dwarf galaxies with LSST will also allow to derive photometric metallicity maps and gain insight into chemical and morphological structure of these stellar systems. 

\begin{figure}
  \epsscale{1.2}
  \plotone{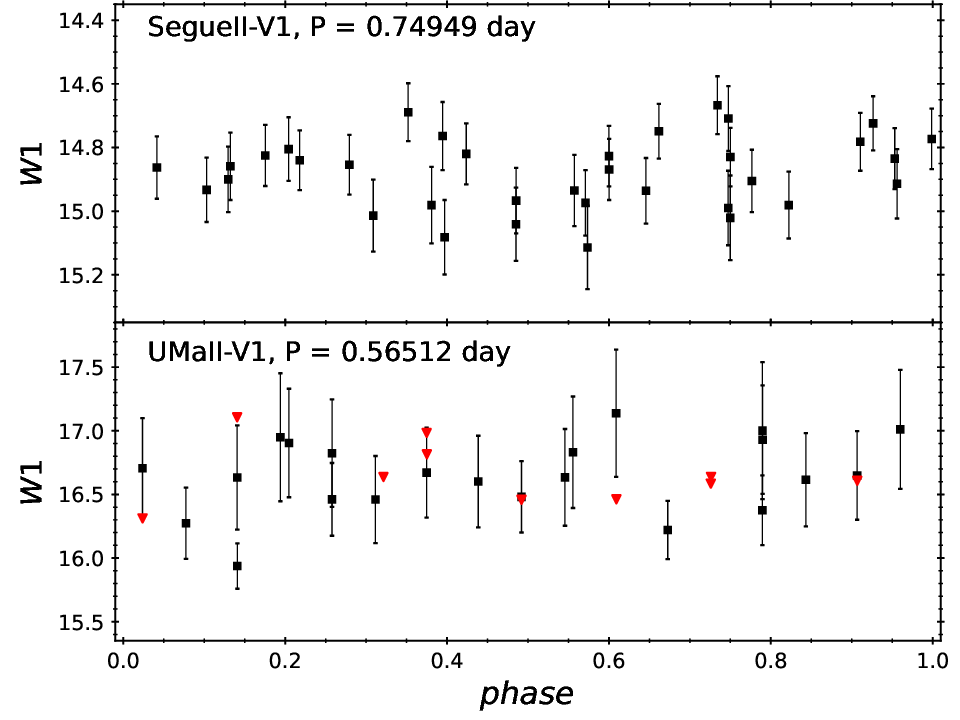}
  \caption{The $W1$-band light curves for SEGUE II-V1 and UMaII-V1 extracted from the ALLWISE \citep{allwise} database, which is composed of data collected from the WISE \citep{wright2010} and the NEOWISE \citep{mainzer2011} mission. The red triangles indicate data-points with ``null'' entries in the error bar. Both light curves were folded with periods adopted from Gaia DR3 (see Section \ref{sec2}).}
  \label{fig_wise}
\end{figure}

Finally, it is worth to remind that the $gr$-band and the $VI$-band metallicity-Fourier parameter(s)-period relations used in Section \ref{sec3} are in the Pan-STARRS1 and the Johnson-Kron-Cousins photometric systems, respectively. Therefore, the $gri$-band light curves based on the LSST observations need be photometrically transformed to the $grVI$-band, as well as the Gaia $G$-band, by properly taking the color-terms into account \citep[as discussed in][]{ngeow2022}. We expect such photometric transformations would be available after the first-light of LSST, and hence it will be straight forward to apply our proposed approach to the LSST light curve data for RR Lyrae.

\acknowledgments

We thank the useful discussions and comments from an anonymous referee to improve the manuscript. CCN is thankful for funding from the National Science and Technology Council (NSTC, Taiwan) under the contract 109-2112-M-008-014-MY3. AB acknowledges funding from the European Union's Horizon 2020 research and innovation program under the Marie Skłodowska-Curie grant agreement No. 886298. This publication has made use of data collected at Lulin Observatory, partly supported by NSTC grant 109-2112-M-008-001. This research has made use of the SIMBAD database and the VizieR catalogue access tool, operated at CDS, Strasbourg, France. This research made use of Astropy,\footnote{\url{http://www.astropy.org}} a community-developed core Python package for Astronomy \citep{astropy2013, astropy2018, astropy2022}. 

This work has made use of data from the European Space Agency (ESA) mission {\it Gaia} (\url{https://www.cosmos.esa.int/gaia}), processed by the {\it Gaia} Data Processing and Analysis Consortium (DPAC,
\url{https://www.cosmos.esa.int/web/gaia/dpac/consortium}). Funding for the DPAC has been provided by national institutions, in particular the institutions participating in the {\it Gaia} Multilateral Agreement.

The Pan-STARRS1 Surveys (PS1) and the PS1 public science archive have been made possible through contributions by the Institute for Astronomy, the University of Hawaii, the Pan-STARRS Project Office, the Max-Planck Society and its participating institutes, the Max Planck Institute for Astronomy, Heidelberg and the Max Planck Institute for Extraterrestrial Physics, Garching, The Johns Hopkins University, Durham University, the University of Edinburgh, the Queen's University Belfast, the Harvard-Smithsonian Center for Astrophysics, the Las Cumbres Observatory Global Telescope Network Incorporated, the National Central University of Taiwan, the Space Telescope Science Institute, the National Aeronautics and Space Administration under Grant No. NNX08AR22G issued through the Planetary Science Division of the NASA Science Mission Directorate, the National Science Foundation Grant No. AST-1238877, the University of Maryland, Eotvos Lorand University (ELTE), the Los Alamos National Laboratory, and the Gordon and Betty Moore Foundation.

This publication makes use of data products from the Wide-field Infrared Survey Explorer, which is a joint project of the University of California, Los Angeles, and the Jet Propulsion Laboratory/California Institute of Technology, and NEOWISE, which is a project of the Jet Propulsion Laboratory/California Institute of Technology. WISE and NEOWISE are funded by the National Aeronautics and Space Administration.

\facility{LO:1m, Gaia, PS1}

\software{{\tt astropy} \citep{astropy2013,astropy2018,astropy2022}, {\tt PSFEx} \citep{bertin2011} , {\tt SCAMP} \citep{scamp2006}, {\tt Source-Extractor} \citep{bertin1996}}

\appendix

\section{Details on Photometric Calibration} 

Same as in \citet{ngeow2022}, photometric calibrations on the LOT reduced images were done using the Pan-STARRS1 (PS1) Data Release 1 photometric data \citep{chambers2016,flewelling2020} as reference catalogs. The selection criteria of the PS1 reference stars are identical to \citet[][whenever applicable]{ngeow2022}, and will not be repeated here. In the $gr$-band, the photometric calibration was performed via the following equations:

\begin{eqnarray}
  g^{PS1} - g^{\mathrm{instr}} & = &  ZP_g + C_g(g^{PS1}-r^{PS1}), \label{eqn_g}\\
  r^{PS1} - r^{\mathrm{instr}} & = &  ZP_r + C_r(g^{PS1}-r^{PS1}). \label{eqn_r}
\end{eqnarray}

\noindent These equations also allow the $(g^{PS1}-r^{PS1})$ colors to be measured via

\begin{eqnarray}
  (g^{PS1}-r^{PS1}) & = & \frac{ZP_g-ZP_r + (g^{\mathrm{instr}} - r^{\mathrm{instr}})}{1-C_g+C_r}. \label{eqn_gr}
\end{eqnarray}

\noindent For the $V$-band, we first transformed the $gr$-band photometry in the PS1 reference stars catalog to the $V$-band using the transformation equation provided by \citet{tonry2012}, followed by fitting the reference stars with the following equation:

\begin{eqnarray}
  V - v^{\mathrm{instr}} & = &  ZP_V + C_V(g^{PS1}-r^{PS1}). \label{eqn_v}
\end{eqnarray}

\noindent Calibration for the $i$-band data required a further treatment, this is because we would like to calibrate the $i$-band photometry to the Johnson-Cousin $I$-band magnitudes. Combining $i^{PS1} - i^{\mathrm{instr}}  =   ZP_i + C_i(g^{PS1}-r^{PS1})$ with the \citet{tonry2012} transformation, the $i$-band data can be calibrated using the following equation:

\begin{eqnarray}
  I - i^{\mathrm{instr}} & = &  (ZP_i - 0.367) + (C_i - 0.149) \times (g^{PS1}-r^{PS1}). \label{eqn_i}
\end{eqnarray}

\noindent In equations \ref{eqn_g} to \ref{eqn_i}, $\{g, r, v, i\}^{\mathrm{instr}}$ are instrumental magnitudes measured with $MAG\_PSF$ implemented in the {\tt Source-Extractor} package \citep[together with the PSF models derived from the {\tt PSFEx} package,][]{bertin2011}, while $g^{PS1}$ and $r^{PS1}$ are magnitudes taken from the PS1 reference catalog. After solving the coefficients $ZP$ and $C$ in these equations, we then apply them to the RR Lyrae by first obtaining the $(g^{PS1}-r^{PS1})$ colors using equation \ref{eqn_gr}, and subsequently calibrate the $grVI$-band photometry from the same set of $Vgri$ images.

\section{The Light-Curves in the WISE-Filter} \label{appxb}

Besides the $V$-band relation, \citet{mullen2021} have also derived a similar relation in the infrared WISE filter. However, the $W1$-band light curves for both RR Lyrae, extracted from the ALLWISE database, do not show a RR Lyrae-like light curve in the $W1$-band, rather the extracted light curves are quite scatter with large error bars (see Figure \ref{fig_wise}). As a result, we omitted the $W1$-band $[\mathrm{Fe/H}]$-$\phi_{31}$-$P$ relation in this work.


\end{document}